\documentclass[11pt,twoside]{article}
\usepackage{./asp2014}
\usepackage{graphicx}

\aspSuppressVolSlug
\resetcounters

\begin{document}

\title{Disk formation in oblate B[e] stars}
\author{Araya, I., Arcos, C., and Cur\'e, M.
\affil{Instituto de F\'isica y Astronom\'ia, Universidad de Valpara\'iso, Valpara\'iso, Chile; \email{ignacio.araya@uv.cl}}}

% This section is for ADS Processing.  There must be one line per author.
\paperauthor{Araya, I}{ignacio.araya@uv.cl}{}{Universidad de Valpara\'iso}{Instituto de F\'isica y Astornom\'ia}{Valpara\'iso}{Valpara\'iso}{Postal Code}{Chile}
\paperauthor{Arcos, C}{catalina.arcos@uv.cl}{}{Universidad de Valpara\'iso}{Instituto de F\'isica y Astornom\'ia}{Valpara\'iso}{Valpara\'iso}{Post	al Code}{Chile}
\paperauthor{Cur\'e, M}{michel.cure@uv.cl}{}{Universidad de Valpara\'iso}{Instituto de F\'isica y Astornom\'ia}{Valpara\'iso}{Valpara\'iso}{Postal Code}{Chile}

\begin{abstract}
We investigate the possible role of line-driven winds in the circumstellar envelope in B[e] stars, mainly
the role of the $\Omega$-slow wind solution, which is characterized by a slower terminal velocity and
higher mass-loss rate, in comparison with the standard (m-CAK) wind solution. In this work, we assume two scenarios: 
1) a spherically symmetric star and 2) a scenario that considers the oblate shape, considering only the oblate correction factor. 
For certain values of the line force parameters (according to previous works), we obtain in both scenarios a density contrast $\gtrsim10^{2}$ between equatorial and polar densities, characterized for a fast polar wind and a slow and denser wind when the $\Omega$-slow wind solution is obtained. All this properties are enhanced when the oblate correction factor is included in our calculations.
\end{abstract}

\section{Introduction}
The so-called B[e] phenomenon \citep{lamers1998} is typically defined by the
presence of an intense Balmer emission B-type spectrum that, in the optical region, 
simultaneously displays emission lines of low-excited permitted and forbidden transitions 
of low-ionized elements (i.e., \ion{Fe}{ii}, [\ion{Fe}{ii}] and [\ion{O}{i}]).
In addition, it exhibits a very strong near/mid-infrared excess.\\

The B[e] phenomenon is related to the physical conditions of the gaseous and
dusty shells, rings, or disks surrounding a hot star rather than with the
properties of the star itself. Hence, the characteristics of the B[e]
phenomenon are found in a great variety of objects, ranging from intermediate
to high-mass stars in pre- and post-main sequence evolution. One group of
these objects are the massive evolved B[e] supergiants (B[e] SGs).
The origin of circumstellar envelopes of B[e] SGs is usually attributed to the
mass lost from the star via dense stellar winds or sudden mass ejections
expected to occur during short-lived phases in the post-main sequence evolution
of the stars. The structure of the circumstellar envelope was described
by \cite{zickgraf1985} who proposed an empirical model,
which consists in a hot and fast line-driven wind in the polar region, and a slow,
much cooler and denser (by a factor of $10^2-10^3$) wind in the equatorial
region. Such a wind structure could be obtained via the rotation induced
bistability mechanism \citep{lamers1991, pelupessy2000},
or via the bistability combined with the $\Omega$-slow wind solution \citep{cure2004} at high rotation rates \citep{cure2005}.\\

In \cite{cure2005} only one model was calculated based in the work of \cite{pelupessy2000}. In this current work we will analyze a grid of line force parameters plus the addition of the oblateness shape of the star and the oblate correction factor and solve the highly non-linear m-CAK equation of motion.

\section{Hydrodynamic Wind Equations}
The hydrodynamic wind equations belonging to the m-CAK theory \citep{castor1975,abbott1982,friend1986,ppk1986} define a stationary 1-D spherical outflow considering in the Equation of Motion (EoM) the radiation coming from a stellar disk, through the finite disk correction factor ($f_{SFD}$), and the line-force parameters given by $\alpha$, $\delta$ and $k$.  These equations are numerically solved for high values of $\Omega$ ($v_{rot}/v_{crit}$) by the code \textsc{Hydwind} already described in \cite{cure2004}. From the high rotational velocities we expect to obtain $\Omega$-slow solutions instead of the fast (standard m-CAK model) solutions.\\

When the star become oblate due to the rapid rotation, the $f_{SFD}$ should be replaced by the oblate finite disk correction factor ($f_{OFD}$) which is described by \cite{pelupessy2000}. 
The method described by \cite{araya2011} is implemented in order to solve the m-CAK EoM accounting  the oblate distortion of the star. The $f_{OFD}$ can be obtained by an approximate expression via a sixth order polynomial interpolation $Q(u)$ (which is in terms of $u=-R_{*}/r$),

\begin{equation}
\label{poly-fit}
f_{OFD}=Q(u)\,f_{SFD}.
\end{equation}

\noindent This approximation allows to solve in a simpler way the oblate correction factor and therefore the EoM. For comparison, $Q(u)$ and the ratio $f_{OFD}/f_{SFD}$ are shown in the left panel of the Figure~\ref{figure1}. The right panel shows the good agreement between the oblate correction factors calculated numerically by the exact equation and our approximation thanks to the fit of $Q(u)$.

\articlefiguretwo{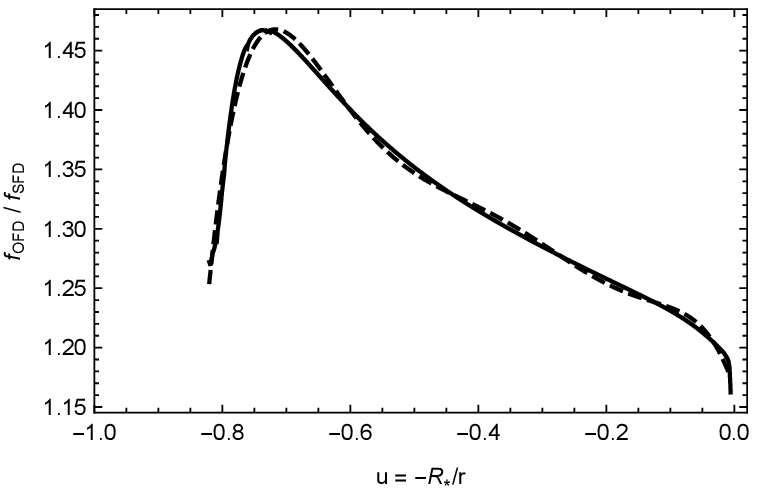}{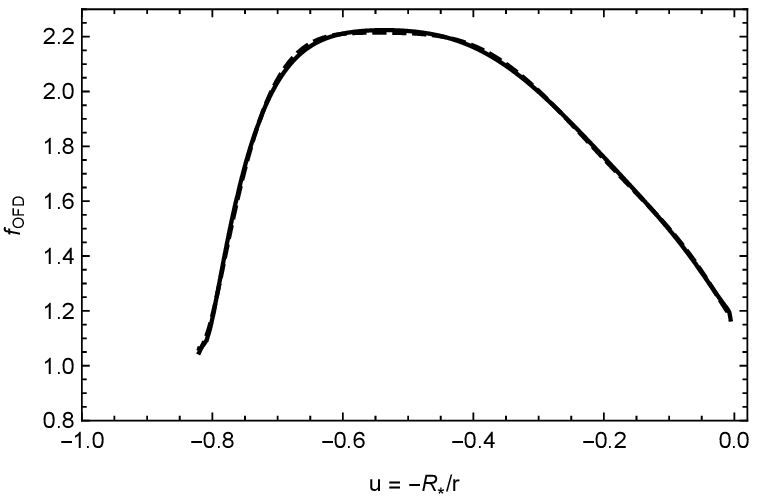}{figure1}{\emph{Left:} Ratio $f_{OFD}/f_{SFD}$ (solid line) and the polynomial interpolation $Q(u)$ (dashed line) for a wind model calculated at the equator and rotating at $\Omega=0.90$.  \emph{Right:} Oblate correction factors calculated by an exact way (solid line) and by our approximate method (dashed line).}

\section{Results}

The hydrodynamic equations are solved for a rotating B[e] supergiant star ($T_{eff}=25\,000[K]$, $M_{*}=17.5\,[M_{\odot}]$ and $L_{*}=10^{5}\,[L_{\odot}]$) considering the $f_{SFD}$ and $f_{OFD}$ for spherical and oblate scenarios, respectively. Note that in our calculations we do not include gravity darkening. Accounting for the line force parameters used by \cite{cure2005} and \cite{pelupessy2000} for this same stellar model, we calculate the equatorial and polar wind solutions for values of $\alpha=0.45,0.50,0.55,0.60,0.65$ and $k=0.06$, and also $k=0.57$ for equatorial models. The value of $\delta$ is equal to $0.0$ in all our cases. A total of 50 models for each scenario (spherical and oblate) were calculated, but only 36 spherical models had a physical solution.  All the models with physical solution have fast solutions for $\Omega=0.6$. For the other cases, we obtained $\Omega$-slow solutions, only in some few cases the models calculated with $\Omega=0.7$ and $0.8$ led fast solutions. \\

\begin{landscape}
\begin{figure}
\includegraphics[scale=0.75]{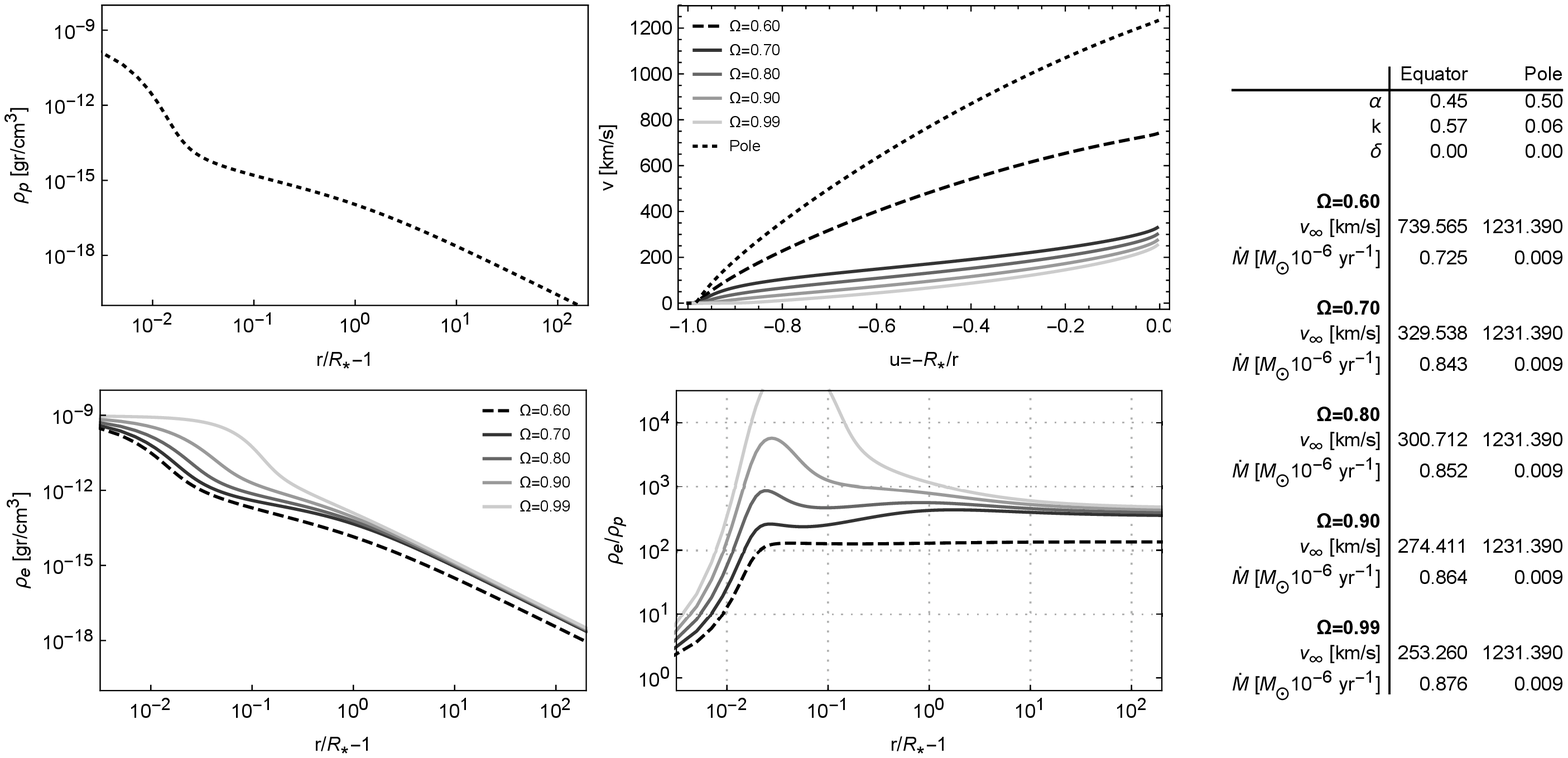}
\vspace*{1.0cm}
\caption{Equatorial and polar hydrodynamic wind solutions for a spherical model considering four rotational velocities. Solid and dashed lines correspond to the equatorial $\Omega$-slow and fast solutions, respectively, while the dotted lines depict the polar wind solutions. Top left and bottom left panels correspond to polar and equatorial densities, respectively. Top right panel shows equatorial and polar velocity profiles. Bottom right panel represents the density contrast between equatorial and polar densities. The table contains the line force and wind parameters for each case.\label{figure2}}
\end{figure}
\end{landscape}

\begin{landscape}
\begin{figure}
\includegraphics[scale=0.75]{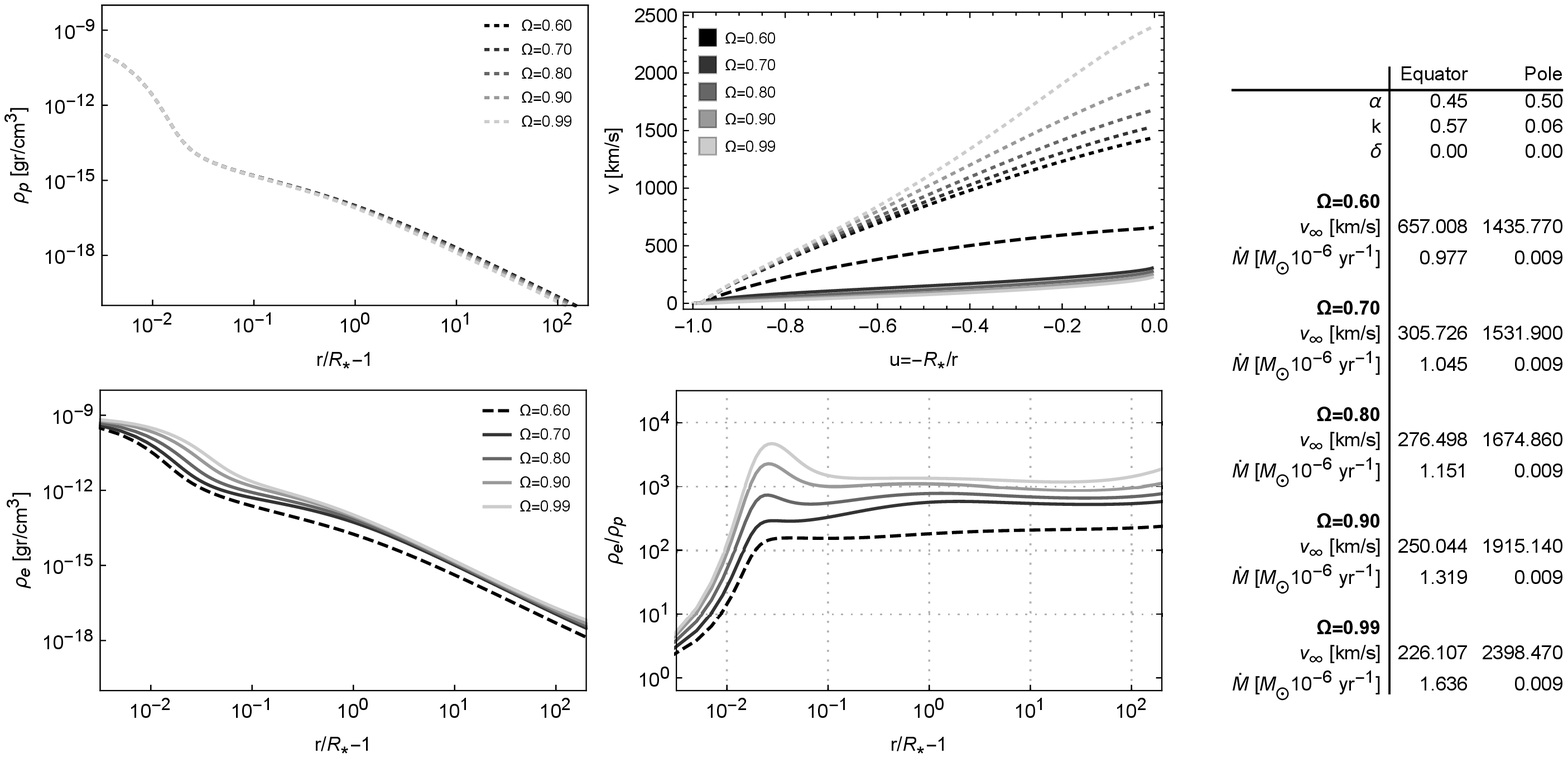}
\vspace*{1.0cm}
\caption{Similar to Figure \ref{figure2}, but with hydrodynamic wind solutions for an oblate model. \label{figure3}}
\end{figure}
\end{landscape}

Figures~\ref{figure2} and ~\ref{figure3} illustrate an example of the results obtained for one set of line force parameters considering spherical symmetry and the oblate shape of the star, respectively. We noted that the density contrast is highly affected by the line force parameters $\alpha$ and $k$. Moreover, for models with $k = 0.57$ at the equator and $k = 0.06$ at the pole, we obtained the highest densities contrast. Note that these values of $k$ are based on the work of  \cite{pelupessy2000}. In the case of $\alpha$, higher values of this parameter produce higher mass loss rates, then the density contrast increments its value when we increase the difference between the $\alpha$ used at the equator and the pole. \\

Comparing both scenarios, the terminal velocities for oblate models have lower values than the spherical models, while the mass loss rates are higher for oblate cases, but only when we use $k=0.57$. Spherical models with $k=0.06$ present a higher mass loss rate. The models considering the oblateness of the star show higher densities contrast, in comparison to the spherical models, when the equatorial models are calculated with $k=0.57$. On the contrary, the spherical models with $k=0.06$ show higher densities contrast than the oblate models.  

\section{Conclusions and Future Work}
We obtained the hydrodynamic solutions of the equatorial and polar wind of a B[e] star for a spherical and an oblate model with rotational velocities over $\Omega=0.60$.  In both models, the terminal velocity of the equatorial wind decrease as the rotational velocity of the star increases, but the oblate models lead  lower values of the equatorial terminal velocity in comparison with the spherical models. The polar terminal velocities are higher for oblate cases than the spherical models due to the large radiation from the stellar surface. The density contrast between equatorial and polar densities depend mostly of the line force parameters. The highest densities contrast ($\gtrsim10^{2}$) are obtained when the $\Omega$-slow solution is obtained at the equator and by models using $k$ values equal to $0.57$ and $0.06$ at the equator and the pole, respectively. In the case of the models considering the oblate shape of the star the density contrast is enhanced when we use this same set of parameters, $k=0.57$ (equator) and $k=0.06$ (pole).\\

The next step in our study is test if our predictions are in agreement with observations. The spectrum of B[e] supergiants are typically represented by the P Cygni profiles of UV and Balmer lines. The development of radiative transfer code is necessary to obtain the spectrum and spectral energy distribution of these massive star. \textsc{Hdust} \citep{carciofi2006} is a non-LTE Monte-Carlo radiative transfer code that take in account the distortion of the central star and the gravity darkening effect due to the fast rotation. The code is able to reproduce Balmer lines, spectral energy distribution, polarization and images from a wind or gas/dust disk models, using different options by model. One of these options -- under construction -- is to read an external file containing the density and velocity profiles of the wind. Once the options is available, first we will procedure to compare theoretical profiles of Balmer lines in the optical (i.e, H$\alpha$) and in the NIR (Br$\alpha$) and we will compare both results, spherical and oblate wind. Then, we will take observations from some high rotators ($\Omega \sim$ 0.70) B[e] stars presented in \cite{zickgraf1986} in order to reproduce the observables.

\acknowledgements
The authors thanks the support of the Centro de Astrof\'isica de Valpara\'iso.  I. A. acknowledges support from Fondo Institucional de Becas FIB-UV and Gemini-Conicyt 32120033. M. C. acknowledge and thanks the organizers of this congress for making possible his attendance. C. A. acknowledges partial support from FONDECYT Iniciacion throught grant 11130702, CONICYT Capital Humano Avanzado project N 7912010046, Gemini-CONICYT 32120033 and BECAS DE DOCTORADO NACIONAL CONICYT 2016.

%\bibliography{editor}  % For BibTex
\bibliography{citas}

\end{document}